\documentclass{article}

\usepackage[]{graphicx}

\newcommand{\be}{\begin{equation}}
\newcommand{\ee}{\end{equation}}

\begin{document}


\title{Weightlessness of photons: A quantum effect}         
\author{Ari Brynjolfsson \footnote{Corresponding author: aribrynjolfsson@comcast.net}}

\date{\centering{Applied Radiation Industries, 7 Bridle Path, Wayland, MA 01778, USA}}          

\maketitle

\begin{abstract}  Contrary to general belief, the Fraunhofer lines have been found to be plasma redshifted and not gravitationally redshifted, when observed on Earth.  Quantum mechanical effects cause the photons' gravitational redshift to be reversed as the photons move from the Sun to the Earth.  The designs of the experiments, which were thought to have proven the gravitational redshift of photons, are all in the domain of classical physics, and make it impossible to detect the reversal of the gravitational redshifts.  The solar redshift experiments, however, are in the domain of quantum mechanics; and the reversal of the redshift is easily detected, when the plasma redshift is taken into account.  The photons are found to be weightless relative to a local observer, but repelled relative to a distant observer.  The weightlessness of the photons in the gravitational field relative to a local observer is inconsistent with Einstein's equivalence principle.  This together with the plasma redshift has profound consequences for the cosmological perspectives.  This article gives a theoretical explanation of the observed phenomena, proper interpretation of the many gravitational redshift experiments, and an understanding of how we missed observing the reversal of photons' gravitational redshift.  The present analysis indicates that although the photons are weightless in a local system of reference, the experimental evidence indicates that quasi-static electromagnetic fields are not weightless, but adhere to the principle of equivalence.

\end{abstract}

\noindent  \textbf{Keywords:} General relativity and gravitation, quantum gravity, gravitational theories, physics of black holes, cosmology, redshift, solar redshift, gravitational redshift.

\noindent  \textbf{PACS:} 52.25.Os, 52.40.-w, 97.10.Ex, 04.60.-m, 98.80.Es, 98.70.Vc


\tableofcontents

\makeatletter	   
\renewcommand{\ps@plain}{
     \renewcommand{\@oddhead}{\textit{Ari Brynjolfsson: Weightlessness of photons: A quantum effect}\hfil\textrm{\thepage}}%
     \renewcommand{\@evenhead}{\@oddhead}
     \renewcommand{\@oddfoot}{}
     \renewcommand{\@evenfoot}{\@oddfoot}}
\makeatother     

\pagestyle{plain}


\vspace{14mm}

\section{Introduction}

The discrepancy between the observed and the expected gravitational redshifts in the Sun are often assumed to be caused by Doppler shifts in the line forming elements.  An elaborate system of currents, which was surmised for explaining the discrepancy, has lead to contradictions with observations.  The recently discovered {\it{plasma redshift}} of photons explains the observed redshift of solar Fraunhofer lines [1] (see sections 5.6.1 - 5.6.3 of that source).   The solar Fraunhofer lines are not gravitationally redshifted when observed on the Earth, but plasma redshifted as they penetrate the hot corona of the Sun.  The photons are gravitationally redshifted when emitted in the photosphere, but during their travel from the Sun to the Earth the gravitational redshift is reversed.   In the following, we give a theoretical explanation of these experimental findings.   The reversal of the gravitational redshift is a quantum mechanical effect; and the explanation leads to a modification of the {\it{classical}} general theory of relativity (GTR).  We call this modification the {\it{quantum mechanically modified}} GTR, or simply the modified GTR.  The experiments that have been assumed to confirm the gravitational redshift, such as the experiments by Pound and Rebka Jr. [2,\,3] and Pound and Snider [4,\,5], Vessot et al. [6], and Krisher et al. [7] are all in the domain of classical physics or classical GTR, while the solar redshift experiments are in the domain of quantum mechanically modified GTR when the lines are observed on Earth.

\indent  When Einstein developed the classical GTR from the special theory of relativity, he deduced that a photon when emitted from an atom in the Sun would be gravitationally redshifted [8].  In addition, he assumed that when the gravitationally redshifted photons in the Sun moved from the Sun to the Earth, the photons frequency would not change.  In his arguments for this, Einstein considered that light consists of classical electromagnetic waves; that is, that the wave packet for the photons reached from the Sun to the Earth and beyond.  Therefore, when we compare the frequency of a photon emitted from an atom in the Sun with the frequency of a photon emitted from a corresponding atom on Earth, we should find the solar photons to be gravitationally redshifted.

\indent  The {\it{equivalence principle,}} is based on many experiments, including those of von E\"{o}tv\"{o}s [9], and later by Zeeman [10], Dicke [11], Adelberger et al. [12] and Su et al. [13] who all failed to detect any difference between the inertial and gravitational mass.  These experiments indicate that the inertial mass, $m_i ,$ is equal to gravitational mass $m_g {\rm{.}}~$  In the special theory of relativity, Einstein showed [14,\,15] that the total energy, including the kinetic energy, of a particle can be equated with its inertial mass through $E = m_{i} c^{2}.~$  More generally, it can be proven that an inertial mass, $m_i ,$ can be associated with energy such that $E = m_i c^2 ,$ (see for example M{\o}ller [16] and in particular sections 3.5 to 3.7 of that source).  Based on these findings, Einstein generalized and assumed that the gravitational field attracted all forms of energy, $E,$ including the photon energy, as if $E=h\nu$ were a material particle with $m_g = m_i = h\nu/c^2 {\rm{.}}~$  This assumption, {\it{the equivalence principle},} states that for any form of energy, the gravitational mass $m_g $ is equal to the inertial mass $m_i {\rm{.}}~$

\indent  In classical physics, Einstein's assumption that the photons frequency is a constant of motion as the photon moves from Sun to the Earth appears reasonable; and his assumption that the inertial mass is equivalent to the gravitational mass, {\it{the principle of equivalence},} also appears reasonable,.  His logical deductions are of course correct.  Therefore, it has been generally assumed that his gravitational redshift theory is correct.  But as Einstein made clear, just because the assumptions appear reasonable does not mean that they are correct.  (Aristotle felt it reasonable that: the heavier things are, the faster they fall).  The theory must be tested by appropriate experiments.  Only nature can decide for us if the assumptions are correct (Galileo).

\indent  For an atomic particle, we set $ H = E + V ,$ where $H$ is the total energy or the Hamiltonian, $E$ the kinetic energy and $V$ the potential energy.  When we lift the atomic particle very slowly out of the solar gravitational field, its kinetic energy $E$ remains insignificant, while the Hamiltonian $H$ increases with the gravitational potential energy $V .~$   The gravitational redshift of all the energy levels of the atom in the Sun are then reversed (blue shifted) as we move the atom to the Earth.  Arriving on the Earth, the atom becomes identical to corresponding atoms on Earth.  The blue shifts of all the energy levels during the travel to Earth reverse exactly the gravitational redshifts of the atom's energy levels in the Sun.  The energy, which we transferred to the atom when lifting it from the Sun to the Earth, corresponds to the potential difference.  The energy we transfer to the atom when lifting it causes the blue shifts of all nuclear and atomic energy levels.

\indent  Einstein considered that the photons behave differently.  We cannot bring the photon slowly to the Earth.  Einstein assumed that when the photon particle moves out of the gravitational field, we have that the Hamiltonian $H = E + V = h\nu $ is a constant of motion.  Relative to a distant observer, Einstein assumed that the value of the kinetic energy $E$ decreases as the potential energy $V$ increases such that $H = h\nu $ is a constant of motion, as the photon moves outwards from the Sun.  The energy changes are analogous to that of a stone thrown upwards from Earth.  This is consistent with Einstein's {\it{equivalence principle,}} which assumes that all forms of energy, including that of the photon, are attracted by the gravitational field.  However, based on the solar redshift experiments [1] and the theoretical analyses in section 3, we find these assumptions incorrect.

\indent   The plasma redshift, which is based on well established physics, explains well the observed redshift of the solar Fraunhofer lines.  Observations detect no gravitational redshift [1] (see section 5.6.2 and 5.6.3 of that source).  Einstein's assumptions, thus, lead to contradictions with observations.

\indent  Einstein's assumption that {\it{equally many waves arrive on Earth as leave the Sun}} appears reasonable in framework of classical physics.  But a physicist A, who is a proponent of quantum mechanics, finds Einstein's assumptions impermissible.  Physicist A considers light as consisting of photons with limited length.  He sees no requisite from theory or experiments for the assumption that equally many waves arrive on the Earth as leave the Sun, and he sees no need from experiments or theory to assume that photons are attracted like material particles in the gravitational field.

\indent  Von E\"{o}tv\"{o}s [9] and many others carrying out similar experiments did not work with photons.  But other experiments, for example, the experiments by Pound and Repka, Jr. [2,\, 3], and Pound and Snider [4,\, 5] are generally believed to show that gravitational field attracts photons like material particles.  These experiments are believed to confirm that the {\it{principle of equivalence}} applies to photons.  However, in the following, we show that the interpretation of these experiments is invalid.  The experiments are in the domain of classical physics and can't detect quantum mechanical effects.  These experiments, which have been surmised to confirm the gravitational redshift of photons and {\it{the principle of equivalence for photons},} are in fact insensitive to attraction, weightlessness, or repulsion of the photons.     

\indent  The 14.4 keV photons used in the experiments by Pound and Repka, Jr. [2,\, 3], and Pound and Snider [4,\, 5] did not get enough time, only about $7.5\cdot 10^{-8}~ {\rm{s}} ,$ to change their frequency, as the photons traveled the short distance of only 22.5 m between emitter and absorber in the weak gravitational potential of the Earth.  The minimum time required for changing a photon's frequency is about $1.9 \cdot 10^{-5}~ {\rm{s}} .~$ This minimum time may be derived from the transition probability or the uncertainty relation in quantum mechanics.  The difference in gravitational potential of the 14.4 keV photons at the position of the emitter and the absorber in the experiments [2-5] is
\be
\Delta E = \left( {{{h\nu } \mathord{\left/{\vphantom {{h\nu } {c^2 }}} \right.
 \kern-\nulldelimiterspace} {c^2 }}} \right)981 \cdot 2250 = 5.67 \cdot
 10^{ - 23}\; \; {\rm{ erg,}}
\ee
\noindent  where in the local system of reference the photon's kinetic energy, $E,$ is equal to  $h$\^{$\!\!\!\nu$} $= h\nu /(1+2\chi/c^2),$ which is equal to the difference between the two energy levels $E_2$ and $E_1$ in the nucleus of $_{26}{\rm{Fe}}^{57} $-isotope.  \^{$\!\!\!\nu$} is the standard frequency (the frequency measured by a standard clock, or the frequency measured spectroscopically with the spectroscope placed near the source).  $\nu$ is the coordinate frequency, and $\chi$ is the scalar gravitational potential.  In the classical theory the gravitational mass, $m_g ,$ of the photon is equal to the inertial mass, $ m_i = {{{h\nu } \mathord{/{\vphantom {{h\nu } {c^2 }}} \kern-\nulldelimiterspace} {c^2 }}}{\rm{.}}~$ The factor $981 ~{\rm{cm\,s}}^{-2}$ is the gravitational acceleration acting on the photon, and the factor 2250 cm is the height difference.

\indent  From both the uncertainty relation and the transition theory in quantum mechanics, the minimum time, $\Delta t ,$ required for observing the transition between the two states is approximately
\be
\Delta t \geq {h \mathord{\left/{\vphantom {h {\left( {2\pi \Delta E} \right)}}} \right.
 \kern-\nulldelimiterspace} {\left( {2\pi \Delta E} \right)}} = 1.9 \cdot 10^{ - 5} \;\; {\rm{ s,}}
\ee
\noindent  where $h$ is Planck's constant.  The length of the 14.41 keV photon is about 
$L\approx 2\pi c\tau = 270~ \rm{m},$ where $\tau =143~ \rm{ns}$ is the life time of the 14.41 keV transition in the nucleus.  During the travel from the emitter to absorber, the photons would usually experience smaller potential difference.  Therefore, it is impossible for the photons to adjust to the new potential or to change frequency during their short travel time of $7.5 \cdot 10^{-8}~ {\rm{s}} $ from the emitter to the absorber in these experiments.  On the other hand, the atoms and the nuclei in the emitter and absorber had plenty of time to adjust to the gravitational potentials.  Every transition in quantum mechanics takes time.  Photons require time to change from one state to another, even if the states overlap and are continuous.  In the experiments [2-7] the photons had no chance to adjust to the gravitational potential while they moved from the emitter to the absorber.  To the physicist A, who is familiar with quantum mechanics, these experiments are inconclusive with respect to gravitational redshift theory, because the experiments did not make it possible to detect a weightlessness, a repulsion, or an attraction of a photon by the gravitational field.  

\indent In the rocket experiments by Vessot et al.~[6], the maser photons are too long.  Also the potential difference between emitter and detector is too small.  In the space experiments by Krisher et al.~[7], the laser photons (signals) used in the experiment are too long.  They reach from the Sun to beyond Saturn.  These experiments, which were assumed to have proven the gravitational redshift, are insensitive to the relevant quantum mechanical effects in the interactions of photons with the gravitational field.

\indent Quantum mechanics, in accordance with Bohr's correspondence principle, approaches the classical mechanics in the limit of long-wavelength experiments, and in the limit of slow variation in the gravitational field.  A correct quantum mechanically modified GTR does not contradict, therefore, the classical physics experiments, like the experiments by Pound and Rebka [2-3], Pound and Snider [4-5], rocket experiments by Vessot et al.~[6], and the space experiments by Krisher et al.~[7], because the experiments are inconclusive and do not make it possible to detect a weightlessness, a repulsion, or an attraction of the photon.

\indent  In the gravitational deflection experiments analyzed by Riveros and Vucetich [17], and in the experiments on the time delay of radar echoes by Shapiro et al.~[18], it is important to realize that 50\% of the measured effect is due to variation in the speed of light in the gravitational field, and 50\% is due to warping of space.  The changes in the speed of light and the warping of space are caused by the mass of the Sun (the star or the galaxy), and are independent of weightlessness, repulsion, or an attraction of the photon.  The deflection and the time delay in the above-mentioned experiments are independent of frequency or any frequency changes during time of flight.

\indent Einstein usually used {\it{coordinate frequency}} for describing the phenomena.  Many scientists usually use {\it{standard frequency}} instead, often without stating it clearly.  Using standard frequency, they explain that the photons gravitational redshift is caused by photons loss of kinetic energy, $E,$ as the photons climb the gravitational potential when traveling from the Sun to the Earth.  Use of standard frequency and standard time corresponds to using local clocks at each point P along the track; that is, different clocks at each point.  The reference system, thus, changes with the location of the point P.  In contrast, Einstein usually uses one coordinate system and a coordinate frequency and shows that the frequency is proportional to the total energy $h\nu = H = E + V ,$ which is a constant of motion.  $H$ is the Hamiltonian and $V$ the potential energy.  The kinetic energy, $E,$ decreases by the same amount $V$ increases.  We will with Einstein usually use one coordinate system when comparing the frequencies and energies at different points.  We will with Einstein use a reference system of a distant observer using coordinate clocks.  Einstein's theory has it that the energy levels and the frequencies of the atoms on Earth are higher than the energy levels and the frequencies of the same atoms in the Sun, because of the gravitational potential difference.  Einstein assumed that the photon's coordinate frequency and $h\nu = H$ stays constant as the photon moves from Sun to Earth [8].  (See in particular second paragraph after Eq.\,(2a) of that source.)  In fact, a maser clock in the Sun, according to Einstein, would be seen to move at a slower rate than the clocks on the Earth.

\indent The following section 2 serves only to underscore the well-known fact that a quantum mechanical transition to a new energy state is not instantaneous, but takes a finite time.

\indent  Section 3, the main section, consists of 7 subsections 3.1 to 3.7.  These subsections show that the photon coordinate frequency does not need to be a constant of motion when the photon moves through a gravitational field.  Instead, the photon's coordinate frequency could increase when a photon moves away from a gravitating body, provided there is enough time for the frequency to change (see Eqs.\,1 and 2).  This is consistent with solar redshift experiments, which show that the gravitational redshift is reversed as photons move outwards.  To a distant observer, the photons appear to be pushed outwards. 

\indent  As shown by Einstein, a solar photon is gravitationally redshifted when emitted.  But following the emission, the photon's energy increases during its time of flight from the Sun to the Earth such as to reverse its gravitational redshift [1].  The necessary conditions for this frequency change are:

\begin{enumerate}
\item	The photon can be enclosed in a box of finite dimensions.
\item	The dimensions of the box are so large that the integration of the photon field across the surface of this box is approximately zero. 
\item	The box is so small that the metric and the gravitational potentials are approximately uniform (constant) over the photon.
\item	The distance between emission and absorption of the photon is much larger than the dimensions of the box.
\end{enumerate}

\noindent These four conditions are made strict for simplifying the discussion.  A more elaborate theory can subsequently modify these assumptions for making the conditions gradual as we approach the classical limit.  If these assumptions are not fulfilled, we are in the domain of classical physics behavior of the photon, or on the borderline of classical physics behavior.  For example, when the frequencies become so low or the wavelength so long that we cannot enclose the photon in a box with dimensions smaller than the dimensions of the experiment, like in the experiments by Pound et al. [2-5], we cross over into the domain of classical physics.  This limit for the transition from quantum mechanics to classical mechanics is in accordance with Bohr's correspondence principle; that is, in the long wavelength limit the quantum mechanical prediction of an experimental measurement do not contradict the classical physics prediction.  

\indent  If we can enclose the photon in a relatively small box (small compared to variations in the gravitational field and small compared to the distance traveled) the photon frequency is not a constant of motion, but varies with the gravitational potential as it moves in the gravitational field.  This is a quantum mechanical effect, and the experiment is in the domain of the {\it{modified}} GTR.  For those familiar with the transition from quantum mechanics to classical mechanics, this is not a surprising outcome.  However, this small change has tremendous consequences for the classical general theory of relativity, the equivalence principle, the cosmological theories, and our cosmological perspectives, as shown in reference [1], [19], and [20].

\indent  In section 6 we summarize briefly the conclusions.  We underscore that the theoretical analysis is consistent with weightlessness of photons, and consistent with the weight of the classical electromagnetic fields of electrons and nuclei.


\section{Finite transition time}

Quantum mechanical transition theory makes it clear that any transition from one state to another takes time.  The Hamiltonian for two weakly interacting systems 1 and 2 may be given by $H(1,\,2) = H(1) + H(2) + V(1,\,2) ,$ where $V(1,\,2)$ is the interaction operator.  $H(1)$ may correspond to the Hamiltonian for an atom and $H(2)$ to the Hamiltonian for the electromagnetic field.  The probability $P(t)$ that at time $t$ the system is in a certain state may be given by
\be
P\left( t \right) = \left| {A_a \left( t \right)} \right|^2  = \left| {\int {\left\{ {{\rm{exp}}\left( { - {{iEt} \mathord{\left/
 {\vphantom {{iEt} \hbar }} \right.
 \kern-\nulldelimiterspace} \hbar }} \right)} \right\}\;I_a \left( E \right)\, dE} } \right|^2 {\rm{,}} 
\ee
\noindent where 
\be
A_a \left( t \right) = \int {\psi ^ *  \left( {x_1 ,\;x_2  ,\;0} \right)\,\,} \psi \left( {x_1 ,\;x_2,\;t} \right)\,dx_1 dx_2 
\ee
\noindent and where $\psi (x_1,\,x_2,\,0)$ and $\psi (x_1,\,x_2,\,t)$ are the wave functions for the system, and where 
\be
I_a \left( E \right) = \left| {Q_a \left( E \right)} \right|^2 
\ee
\noindent is the energy distribution function in the state $|a \rangle \!=\! \int {Q_a \left( E \right)}\,{\psi   \left( {x_1 ,\;x_2 } \right)}\, dE ,$  and $I_a(E) dE $ the probability that the system remains in the state $|a\rangle $.  Let $I_a (E)$ be given by the energy dispersion function for photons from excited state to a ground state [1] (see in particular Appendix A of that source), where
\be
I_a \left( E \right)\;dE = \frac{1}{{2\pi }}\frac{{\hbar \,\gamma \,dE}}{{\left( {E - E_a } \right)^2  + {{\hbar ^2 \gamma ^2 } \mathord{\left/
 {\vphantom {{\hbar ^2 \gamma ^2 } 4}} \right.
 \kern-\nulldelimiterspace} 4}}} = \frac{1}{{2\pi }}\frac{{\gamma \,d\omega }}{{\left( {\omega  - \omega _a } \right)^2  + {{\gamma ^2 } \mathord{\left/
 {\vphantom {{\gamma ^2 } 4}} \right.
 \kern-\nulldelimiterspace} 4}}}{\rm{.}} 
\ee
\noindent From these relations, we have then that 
\be
\int {\left\{ {{\rm{exp}}\left( {{{ - i\,E\,t} \mathord{\left/
 {\vphantom {{ - i\,E\,t} \hbar }} \right.
 \kern-\nulldelimiterspace} \hbar }} \right)} \right\}} \,I_a \left( E \right)\;dE = {\rm{exp}}\left[ { - i{{E_a \,t} \mathord{\left/
 {\vphantom {{E_a \,t} \hbar }} \right.
 \kern-\nulldelimiterspace} \hbar } - {{\gamma \,t} \mathord{\left/
 {\vphantom {{\gamma \,t} 2}} \right.
 \kern-\nulldelimiterspace} 2}} \right]
\ee
\noindent and that the exponential decay is
\be
P\left( t \right) = {\rm{exp}}\left( { - \gamma \,t} \right){\rm{ = exp}}\left( { - \frac{{\Delta E_w \,t}}{\hbar }} \right) = {\rm{ exp}}\left( { - \frac{t}{\tau }} \right){\rm{,}}
\ee
\noindent We have also that $\gamma = \gamma_u + \gamma_l = 1/\tau,$ where $\gamma_u$ and $\gamma_l$ are the widths of the upper and the lower level, respectively, and $\gamma$ is the width in the decay from the upper to the lower state with a possible emission of a photon [21], and $\tau$ is the lifetime of the excited state.


\section{Photons interactions with the gravitational fields}

In the following calculations, we treat the photon semi-classically.  This simplification prevents us from dealing with creation and annihilation of photons in the gravitational field.  We assume that these effects have only minor effect in the proposed experiments.  Later, the modified theory can be extended to include such effects.  The present section serves merely as a phenomenological description of photons in the gravitational field, and is analogous to the phenomenological description of the falling apple in the classical theory.  It does not deal with the details of how or why the gravitational field interacts with a particle or a photon.


\subsection{The metric tensors in a gravitational field}

The effect of the gravitational field can be determined with help of the metric tensor.  This metric tensor for a permanent gravitational field from a central body is obtained with help of Schwarzschild exterior solution, which in a spherical coordinate system, $ x^i = \left( r, \, \theta, \, \varphi, \, ct \right),$ shows that the metric tensor at the position P is (see Eqs.\,(11.83) and (12.36) of reference [16]) 
\be
g_{ik}  = \left\{ {\begin{array}{*{20}c}
   {\varepsilon ^2 } & 0 & 0 & 0  \\
   0 & {r^2 } & 0 & 0  \\
   0 & 0 & {r^2 {\rm{sin}}^2 \theta } & 0  \\
   0 & 0 & 0 & { - {1 \mathord{\left/
 {\vphantom {1 {\varepsilon ^2 }}} \right.
 \kern-\nulldelimiterspace} {\varepsilon ^2 }}}  \\
\end{array}} \right\} . 
\ee
\noindent In Eq.\,(9), we use the shorthand notation $\varepsilon^2,$ where
\be
\varepsilon ^2  = \frac{1}{{1 + {{2\chi } \mathord{\left/
 {\vphantom {{2\chi } {c^2 }}} \right.
 \kern-\nulldelimiterspace} {c^2 }}}}{\rm{ = }}\frac{1}{{1 - {{2GM} \mathord{\left/
 {\vphantom {{2GM} {\left( {rc^2 } \right)}}} \right.
 \kern-\nulldelimiterspace} {\left( {rc^2 } \right)}}}}{\rm{ = }}\frac{1}{{1 - {{\alpha} \mathord{\left/
 {\vphantom {{\alpha} {r }}} \right.
 \kern-\nulldelimiterspace} {r }}}}, 
\ee
\noindent and where $\chi = - GM/r $  is the gravitational potential, and $G = 6.67\cdot 10^{-8} ~ {\rm{cm}}^3 \, {\rm{s}}^{-2} \, {\rm{g}}^{-1}$  is the Newtonian constant of gravitation, $M$ the mass in gram of the star (Sun), and $r$ the distance in cm from the star's center to the position P.  The value of $ c $ is equal to the speed of light far away from a gravitating body.  $\alpha = 2GM/c^2$ is an often used parameter.  The three components $g_{\iota\, 4} = 0 ,$ where $\iota = 1,~2,~\rm{or}~3 .~$ Therefore, the speed of light at each point is isotropic and equal to $c^* = c/\varepsilon .~$  (See Eq.\,(8.72) of reference [16]).  According to Eqs.\,(10), we have for $r \rightarrow \infty ,$ that $\chi \rightarrow 0$ and $\varepsilon \rightarrow 1.~$  

\indent  A distant observer finds that at a point P close to the star the spatial units are contracted by a factor $1/\varepsilon$ in all directions, and that the time unit is dilated by a factor $\varepsilon {\rm{.}}~$  One meter measuring rod at P close to the Sun appears to a distant observer to be $1/\varepsilon$ meters.  One second on an atomic clock at P close to the Sun appears to a distant observer to be $\varepsilon $ seconds.

\indent  The contravariant of the metric tensor given by Eq.\,(9) is 
\be
g^{ik}  = \left\{ {\begin{array}{*{20}c}
   {1/\varepsilon ^2 } & 0 & 0 & 0  \\
   0 & {1/r^2 } & 0 & 0  \\
   0 & 0 & {1/r^2 {\rm{sin}}^2 \theta } & 0  \\
   0 & 0 & 0 & { - \varepsilon ^2 }  \\
\end{array}} \right\} . 
\ee
\indent The line element for the metric given by Eqs.\,(9) to (11) is
\be
({ds})^2  = \varepsilon^2 \, {dr}^2 ~ + ~ r^2 \left\{ d \theta ^2  + {\rm{sin}^2} \, \theta \, {d\varphi}^2 \right\} - ( {1 /{\varepsilon ^2 }}) \,  c^2 \, dt^2 . 
\ee
\indent  At a point P where the gravitational field is weak, we may introduce other cordinate systems.  For example, for
\be
\begin{array}{l}
\displaystyle r' = \frac{{1}}{{2}}\left\{{(r^2 - \alpha \, r)}^{1/2} + r - \frac{{\alpha}}{{2}} \right\},\quad {\rm{and}} \\ \displaystyle x = r'\, {\rm{cos}} \, \theta , \quad y = r'\, {\rm{sin}} \, \theta \, {\rm{sin}} \, \varphi , \quad z = r'\, {\rm{sin}} \, \theta \, {\rm{cos}} \, \varphi , 
 \end{array}
\ee
\noindent  the square of the line element at a point P takes the form
\be
ds^2 = \left ( 1 + \frac {{\alpha}}{{r'}} \right ) (dx^2 ~+~ dy^2 ~+~ dz^2 ) - \left ( 1 - \frac {{\alpha}}{{r'}} \right ) c^2 \, dt^2 .
\ee
\noindent  Eqs.\,(13) and (14) may be compared with the Eqs.\,(11.85) (11.90), and (11.91) in reference [16].  For small values of $\alpha /r' , $  Eq.\,(14) can be approximated by 
\be
ds^2 = \varepsilon^2 (dx^2 ~+~ dy^2 ~+~ dz^2 ) - ( {1 /{\varepsilon ^2 }}) \, c^2 \, dt^2 .
\ee
\noindent  In the close surroundings of a point P where ${{2GM} \mathord{\left/ {\vphantom {{2GM} {\left( {rc^2 } \right)}}} \right. \kern-\nulldelimiterspace} {\left( {rc^2 } \right)}} << 1 ,$ we can then use a Cartesian coordinate system with the metric given by 
\be
g_{ik}  = \left\{ {\begin{array}{*{20}c}
   {\varepsilon ^2 } & 0 & 0 & 0  \\
   0 & {\varepsilon ^2  } & 0 & 0  \\
   0 & 0 & {\varepsilon ^2  } & 0  \\
   0 & 0 & 0 & { - {1 \mathord{\left/
 {\vphantom {1 {\varepsilon ^2 }}} \right.
 \kern-\nulldelimiterspace} {\varepsilon ^2 }}}  \\
\end{array}} \right\}, 
\ee
\noindent  and
\be
g^{ik}  = \left\{ {\begin{array}{*{20}c}
   {1/\varepsilon ^2 } & 0 & 0 & 0  \\
   0 & {1/\varepsilon ^2 } & 0 & 0  \\
   0 & 0 & {1/\varepsilon ^2 } & 0  \\
   0 & 0 & 0 & { - \varepsilon ^2 }  \\
\end{array}} \right\} . 
\ee

\vspace{2mm}


\subsection{Electromagnetic wave at P in a system free of gravitational fields}

Let us now consider an electromagnetic plane wave moving outward at P in the direction of the x-axis in a reference system free of gravitational fields.  The metric tensor is given by Eq.\,(16) for $\varepsilon = 1 .~$  In a metric free of gravitational fields and with the dielectric constant and magnetic permeability equal to one, we have that $\mathbf{\tilde{D}}=\mathbf{\tilde{E}}$ and $\mathbf{\tilde{B}}=\mathbf{\tilde{H}}.~$  The tilde over a quantity indicates that the metric is free of gravitational fields.  The general solution of the electromagnetic fields when using Cartesian system of coordinates centered at P may be described by
\be
\mathbf{\tilde{D}} = \mathbf{\vec{e}}_2 \, \tilde f \left( \tilde{t} - \tilde{x} /c\right)~ \; + ~ \; \mathbf{\vec{e}}_3 \, \tilde g \left( \tilde{t} - \tilde{x} /c \right) ,
\ee
\noindent and
\be
\mathbf{\tilde B} = - \mathbf{\vec e}_2 \, \tilde g\left( \tilde{t} - \tilde{x} /c \right)~ \; + ~ \; \mathbf{\vec e}_3 \, \tilde f \left( \tilde{t} - \tilde{x} / c \right) , 
\ee
\noindent where $\mathbf{\tilde D}$ and $\mathbf{\tilde B}$ are the electric and magnetic field vectors, and 
$\mathbf{\vec e}_1 , ~ \mathbf{\vec e}_2,~\rm{and}~ \mathbf{\vec e}_3$ are three unit vectors such that $\mathbf{\vec e}_{1} \mathbf{\vec e}_{2} = \mathbf{\vec e}_{1} \mathbf{\vec e}_{3} = \mathbf{\vec e}_{2} \mathbf{\vec e}_{3} = 0 \,.$

\indent Without loss of generality, we can select the coordinate axis such that 
$\tilde{g} \left( \tilde{t} - \tilde{x} /c \right) = 0 .~$ We have then that Eqs.\,(18) and (19) can be replaced by
\be
\mathbf{\tilde{D}} = \mathbf{\vec{e}}_2 f \left( \tilde{t} - \tilde{x}/c \right) = \tilde{D}_y \mathbf{\vec e}_2 = \varepsilon \, \tilde{E}_y \mathbf{\vec e}_2 , 
\ee
\noindent and
\be
\mathbf{\tilde{B}} = \mathbf{\vec{e}}_3 {f}\left( \tilde{t} - \tilde{x}/c \right) = \tilde{B}_z \mathbf{\vec{e}}_3 =\varepsilon \, \tilde{H}_z \mathbf{\vec{e}}_3 ,
\ee
\noindent and because at the point P the reference system is free of gravitational field, we have that $\varepsilon = 1$ in the metric tensor.  The electromagnetic energy density is then
\be
\tilde{w} = \frac{{\mathbf{\tilde{E}} \cdot \mathbf{\tilde{D}} ~ + ~ \mathbf{\tilde{H}} \cdot \mathbf{\tilde{B}}}}{{2}} = {f}^2 \left( \tilde{t} - \tilde{x}/c \right) ,
\ee
\noindent  and the electromagnetic energy flux is
\be
\mathbf{\tilde{S}} = c \, \mathbf{\tilde{E}} \times \mathbf{\tilde{H}}= c \, {f}^2 \left( \tilde{t} - \tilde{x}/c \right) \, \mathbf{\vec{e}}_1  = c  \, \tilde w \, \mathbf{\vec{e}}_1 .
\ee

\vspace{2mm}


\subsection{Electromagnetic wave at P in a system with gravitational fields}

As shown by M{\o}ller [16] (see sections 10.9 and 10.10 of that source), the electromagnetic quantities are affected by the gravitational field.  In the present case, the axial vectors $\textbf{B},$ and $\textbf{H}$ dual to an antisymmetric tensor $B_{\iota\kappa}$ and $H_{\iota\kappa}$ would with present notations be given by

\be
B^3 = B^z = \frac{{1}}{{\sqrt{\gamma}}}\, B_{1\,2}, \quad  H_3 =H_z = \sqrt{\gamma}\, H^{1\,2}, \quad {\rm{and}} \quad \varepsilon H^{1\,2} = B^{1\,2} ,
\ee
\noindent where $\gamma = {\left| \gamma_{\iota\kappa} \right|} = \varepsilon^6 $ is the determinant of the three space metric tensor $\gamma_{\iota\kappa} .$

\indent In a gravitational field, the electromagnetic covariant and contravariant field tensors are (see the corresponding Eqs.\,(10.286) to (10.288) in reference [16])
\be
F_{ik} = \left\{ {\begin{array}{*{20}c}
   0 & {\sqrt{\gamma}\,\tilde B^z } & 0 & 0  \\
   { - \sqrt{\gamma}\,\tilde B^z  } & 0 & 0 & {\tilde E_y }  \\
   0 & 0 & 0 & 0  \\
   0 & { - \tilde E_y } & 0 & 0  
\end{array}} \right\}, 
\ee
\noindent and
\be
F^{ik}  = \varepsilon \, \left\{ {\begin{array}{*{20}c}
   0 & {{\tilde H_z} /{\sqrt{\gamma}} } & 0 & 0  \\
   { - {\tilde H_z} /{\sqrt{\gamma}} } & 0 & 0 & { - \tilde D^y }  \\
   0 & 0 & 0 & 0  \\
   0 & {\tilde D^y } & 0 & 0  
\end{array}} \right\},
\ee
\noindent where $\varepsilon$ is given by Eq.\,(10), and $\sqrt{\gamma} = \varepsilon^3.~$  As M{\o}ller has shown,  see Eq.\,(10.293) of reference [16], we have for the three dimensional field vectors that 
\be
\mathbf{\tilde{D}} = \varepsilon \mathbf{\tilde{E}} \quad \rm{and} \quad \mathbf{\tilde{B}} = \varepsilon \mathbf{\tilde{H}}\,,
\ee
\indent The electromagnetic energy-momentum tensor is defined as (see Eq.\,(10.305) of [16])
\be
S_i^k  = F_{il} F^{kl}  - \delta _i^k \frac{{ F_{lm}  F^{lm} }}{4}.
\ee
\noindent As is conventional, the symbol $\delta_i^k  = \delta_{ik}$ is equal to 1 for $i = k,$ and equal to 0 for $i \neq k.$  We have then that electromagnetic energy-momentum tensor $ S_i^k $ in a gravitational field is given by
\be
S_i^k  =\, \varepsilon \, \left\{ {\begin{array}{*{20}c}
   {\frac{{\tilde{H}_z \tilde{B}^z+\tilde{E}_y \tilde{D}^y}}{{2}}} & 0 & 0 & { - \tilde{E}_y \tilde{H}_z/\sqrt{\gamma}}  \\
   0 & {\frac{{\tilde{H}_z \tilde{B}^z-\tilde{E}_y \tilde{D}^y}}{{2}}} & 0 & 0  \\
   0 & 0 & {-\frac{{\tilde{H}_z \tilde{B}^z-\tilde{E}_y \tilde{D}^y}}{{2}}} & 0  \\
   {{\sqrt{\gamma}}\,\tilde{D}^y \tilde{B}^z} & 0 & 0 & { - {\frac{{\tilde{H}_z \tilde{B}^z+\tilde{E}_y \tilde{D}^y}}{{2}}}}  
\end{array}} \right\},
\ee
\noindent or
\be
S_i^k  =  \left\{ {\begin{array}{*{20}c}
   {\tilde w} & 0 & 0 & { - \tilde w /({\sqrt{\gamma}}\,{\varepsilon} )}  \\
   0 & 0 & 0 & 0  \\
   0 & 0 & 0 & 0  \\
   {{\sqrt{\gamma}}\,{\varepsilon} \,\tilde w} & 0 & 0 & { - \tilde w}  
\end{array}} \right\}, 
\ee
\noindent  where from Eqs.\,(24) and (27) we have that $\varepsilon(\tilde{H}_z \tilde{B}^z) = \varepsilon (\tilde{H}^{1\,2} \tilde{B}_{1\,2}) = \tilde{B}_z \tilde{B}^z = {f}^2 \left( \tilde{t} - \tilde{x}/c \right) = \varepsilon \tilde{E}_y \tilde{D}^y ,$ and where $\tilde w$ is the energy density given by Eq.\,(22), which is valid when $\varepsilon = 1 .~$ The metric tensor is given by Eq.\,(16).  The electromagnetic energy-momentum tensor, $S_i^k ,$ given by Eqs.\,(29) or (30), corresponds to Eqs.\,(10.306) to (10.308) by M{\o}ller [16].

\indent  As M{\o}ller has shown, the electromagnetic energy density, $w ,$ the energy flux density $\textbf{S}$, and the momentum density, $\textbf{g}$ are respectively

\be
w = -\frac{{1}}{{\varepsilon}}\, S_4^4 = \frac{{1}}{{\varepsilon}}\,(\varepsilon \, {\frac{{\tilde{H}_z \tilde{B}^z+\tilde{E}_y \tilde{D}^y}}{{2}}})= \frac{{1}}{{\varepsilon}}\, {f}^2 \left( \tilde{t} - \tilde{x}/c \right) = \frac{{\tilde w}}{{\varepsilon}}, 
\ee
\noindent and
\be
\mathbf{S} = c \,(\mathbf{E} \times \mathbf{H}) = c \,(\mathbf{E} \times \mathbf{H})^{\iota} = \frac{{c}}{{\varepsilon}}\, ( -S_4^1 ) = \frac{{c}}{{\varepsilon}}\, (\varepsilon \tilde{E}_y \tilde{H}_z/{\sqrt{\gamma}})  = \frac{{c \, \tilde w \,}}{{\varepsilon^2}} = \frac{{c \, w \,}}{{\varepsilon}}=  c_{\ast} \, w \, ,
\ee
\noindent where we have used that $\varepsilon \tilde{H}_z / \sqrt{\gamma} = \varepsilon \tilde{H}^{1\,2} = {f} \left( \tilde{t} - \tilde{x}/c \right);$ and $\varepsilon \tilde{E}_y = \tilde{D}_y = {f} \left( \tilde{t} - \tilde{x}/c \right);$ and where $c_{\ast} = c/\varepsilon$ is the velocity of light in the gravitational field.
\be
\textbf {g} = \frac {{1}}{{c}}\,(\mathbf{D} \times \mathbf{B}) = \frac {{1}}{{c}}\,(\mathbf{D} \times \mathbf{B})_{\iota} = \frac {{1}}{{c \varepsilon}}\, S_1^4  = \frac {{1}}{{c \varepsilon}}\, (\varepsilon \tilde{D}^y \tilde{B}^z \, {\sqrt{\gamma}}) = \frac{{\tilde{w} }}{{c}} = \frac{{w }}{{c_{\ast}}} ,
\ee
\noindent where we have used that $\tilde{B}^z \, \sqrt{\gamma} = \tilde{B}_{1\,2} = {f} \left( \tilde{t} - \tilde{x}/c \right).~$

\indent  These equations can be compared with Eqs.\,(10.309) to (10.311) of reference [16].  From Eq.\,(31), we see that the energy density decreases proportional to $1/\varepsilon .~$   Eq.\,(32) shows that the energy flux decreases faster, or proportional to  $1/{\varepsilon}^2 ,$ because both the speed of light and the energy density decreases proportional to  $1/\varepsilon .$  From Eq.\,(33), we see that the momentum density is independent of the gravitational field.  Eq.\,(32) will be modified in the following when we consider that the field consists of photons, see Eqs.\,(45) and (55).


\subsection{Photon energy in a system free of gravitational field}

The definitions and calculations above are in accordance with the Einstein's classical GTR, as shown by M{\o}ller in sections 10.9 and 10.10 of [16].  We will in this subsection treat the electromagnetic field as consisting of photons in accordance with the semi-classical theory of physics.  The use of semi-classical theory instead of quantum mechanics illustrates in a simple way the fundamentals in the transition from a classical to a quantum mechanical GTR.

\indent  When a system, for example an atom, free of external forces decays exponentially from an excited state with a lifetime $\tilde{\tau} ,$ a photon is emitted.  A tilde over the character means its value in a system free of gravitational field.  The energy of the photon, $\hbar \tilde{\omega}_0 = h \tilde{\nu}_0 ,$ is equal to the energy difference between the initial and final states of the atom.  When using conventional semi-classical approach in a system free of gravitational fields, we often Fourier analyze the electromagnetic field, $\tilde{f}(\tilde{t}-\tilde{x}/c) = \tilde{D}_y(\tilde{t},\, 0) , = \tilde{D}_y (0,\,0) \,\, {\rm{exp}}\left\{ i \, \tilde{\omega}_0 \, \tilde{t} - \tilde{t}/\!(2 \tilde{\tau} ) \right\},$ at a point P where $\tilde{x}=0 ,$ and where $1/\tilde{\tau}$ is the damping constant or width of the photon.  The dielectric constant and the magnetic permeability are both assumed to be equal to one (vacuum).  The Fourier analysis of the photon field gives 
\be
\tilde{D}_y (\tilde{t},\,0) = \tilde{D}_y (0,\,0) \,\, {\rm{exp}} \left( i \, \tilde{\omega}_0 \, \tilde{t} - \frac{{ \tilde{t} }}{{ 2 \tilde{\tau} }}\right) =  \int_{-\infty}^\infty \frac{{\tilde{D}_y(0,\,0)}}{{\, 2\pi \, \left\{ i(\tilde{\omega} - \tilde{\omega}_0 ) + 1/\!( 2\tilde{\tau} ) \right\} }} \,\, {\rm{exp}} \left( i \, \tilde{\omega} \, \tilde{t} \right) \, d\tilde{\omega} \ ,
\ee
\noindent  where
\be
\frac{{\tilde{D}_y (0,\,0)}}{{\, 2\pi \, \left\{ i ( \tilde{\omega} - \tilde{\omega}_0 ) + 1/\!( 2 \tilde{\tau} ) \right\} }} =  \int_0^\infty \! \frac{{\tilde{D}_y (\tilde{t},\,0) \,}}{{2\pi}} \,\, {\rm{exp}} \left( -i \, \tilde{\omega} \, \tilde{t} \right) \, d\tilde{t} \, ,
\ee
\noindent and a similar expression for the other field quantities.  For example, for $ \tilde{B^z} (\tilde{t},\,0) $  we get
\be
\tilde{B}^z (\tilde{t},\,0) =  \int_{-\infty}^\infty \frac{{\tilde{B}^z (0,\,0)}}{{\, 2\pi  \left\{ i(\tilde{\omega} - \tilde{\omega}_0 ) + 1/\!( 2\tilde{\tau} ) \right\} }} \,\, {\rm{exp}} \left( i \, \tilde{\omega} \, \tilde{t}  \right) \, d \tilde{\omega}\, .
\ee
\indent The frequency distribution of photon's Fourier components form the familiar resonance curve, with the width $\gamma = 1/\tau $ similar to that in Eq.\,(6).  The moduli of the Fourier components of the semi-classical photons have the same spectral distribution as the corresponding quantum mechanical photons; see for example Heitler [22].  In a reference system free of gravitational fields, the photon fields may then be represented by Fourier components with frequencies around $\tilde{\omega}_0 ,$ and with a distribution given by the left side of Eq.\,(35).

\indent  The left side of Eq.\,(34) shows that the photon is spread out in time and therefore also in spatial dimensions, because $\tilde{t} = \tilde{x}/c .~$  The photon is thus not represented by a field at a point, but by a field in a volume of space and time.  This spread in space-time is consistent with the transition theory and the uncertainty principle in quantum mechanics.  It is also of interest to note that the moduli of the electrical and magnetic fields making up the photon reduce with time proportional to ${ \rm{exp} }\,(-\tilde{t}/\tilde{\tau}).~$  Usually, the frequency $\tilde{\omega}_0$ is much larger than $1/\tilde{\tau}.~$  For example, for the $L_\alpha $-line in the hydrogen atom, we have that $\omega = 1.55\cdot 10^{16}~{ \rm{s}^{-1}} ,$ while $1/\tau = 4.7\cdot 10^8~{ \rm{s}^{-1}} .~$ We may then sometimes use the average of the moduli over time $\tilde {T},$ where $\tilde{\tau} \gg \tilde{T} \gg 1/\tilde{\omega} .~$  With this averaging over $\tilde {T},$ we get for the modulus for the energy that
\[
 \frac{{1}}{{ \tilde{T} }} \int_0^{{\tilde{T}}}\!\!\! \tilde{E}_y (\tilde{t},\,0) \, \tilde{D}^z (\tilde{t},\,0)\,d\tilde{t} = \frac{{1}}{{2}} \tilde{E}_y (0,\,0) \, \tilde{D}^z (0,\,0) \, {\rm{exp}} \left( - \frac{{ \tilde{t}}}{{ \tilde{\tau} }}\right) \frac{{d\tilde{t}}}{{ \tilde{\tau} }}, 
\]
\noindent  and similar expressions for the other combinations of the field quantities in the electromagnetic energy-momentum tensor $S_i^k $ in Eqs.\,(29) and (30).  When we integrate over the photon, we get
\be
\begin{array}{l}
 \displaystyle 
 \int_0^{\delta {\tilde{t}}}\! \frac{{\tilde{H}_z (0,\,0) \, \tilde{B}^z (0,\,0)\,+\,\tilde{E}_y (0,\,0) \, \tilde{D}^y (0,\,0)}}{{4}}  \, {\rm{exp}} \left( - \frac{{ \tilde{t}}}{{ \tilde{\tau} }}\right) \frac{{d\tilde{t}}}{{ \tilde{\tau} }}\\
\displaystyle \quad \quad \quad \quad \quad \quad = \int_0^{\delta {\tilde{x}}}\!\!\!\!{\tilde{w}_{0}}\, {\rm{exp}} \left( - \frac{{ \tilde{x}}}{{c \tilde{\tau} }}\right) \frac{{ d{\tilde{x}} }}{{c \tilde{\tau} }} = \, {\tilde{w}} ;
\end{array} 
\ee
\noindent  where ${\tilde{w}}_{0} = {\tilde{w}}(0,\,0)$ and ${\tilde{t}} = {\tilde{x}}/c  .~$  At $\tilde{x}=0 ,$ we have that $d{\tilde{w}}/d\tilde{x} ={\tilde{w}_{0}}/(c\tilde{\tau})$ is the energy density per line element $d\tilde{x} .$  

\indent  The photon's wave packet with energy density $\tilde{w}$ may be multiplied by $c$ and integrated along the x-axis for $\delta \tilde{x}\gg {c \tilde{\tau}}.~$  We normalize this to an emission of one photon per second, which results in a photon's energy flux, $\hbar \tilde{\omega},$ at a point P of emission where $\tilde{x}=0 .~$  In a metric free of gravitational field the photon can then be represented by
\be
 \mathbf{S} = \!\! \int_0^{\delta {\tilde{x}}}\!\!\!\!\! {c \, {\tilde{w}_{0}}} \, {\rm{exp}}\left(-\frac{{ \tilde{x} }}{{c \, \tilde{\tau}}} \right)  \frac{{ d{\tilde{x}} }}{{c \tilde{\tau} }} =  c \, {\tilde{w}_{0}}  = \hbar \, \tilde{\omega}_0 , 
\ee
\noindent where the upper integration limit may be set equal to the length,
$\delta \tilde{x} \gg c\, {\tilde{\tau}} .~$  While  $\delta \tilde{x} \gg c\, {\tilde{\tau}} ,$ it is assumed also in the present calculations that $\delta \tilde{x}$ is so small that the metric is constant over the volume of the photon, and small compared with the dimensions of the experiment.  We often set $\delta \tilde{x} = 2\pi \, c \, \tilde{\tau} .~$  In the solar redshift experiments, we need a slightly longer length, for example, $\delta x \approx 18 \, c \, \tilde{\tau} ,$ which defines the energy in photon's redshift within 0.72\,\% of the solar gravitational redshift, because ${\rm{exp}}(-18) = 1.5\cdot 10^{-8} ,$ while solar redshift is $2.12\cdot 10^{-6} .~$  When we integrate over time $\tilde{t}$, and normalize the energy flux in the classical electromagnetic field to one photon per square centimeter and per second in a system free of gravitational fields, we have also that
\be
\mathbf{S} = \int_0^{n\tilde\tau}\!\! c \, {\tilde{w}_{0}} \,\, {\rm{exp}}\left(-\frac{{ \tilde{t} }}{{\tilde{\tau}}}\right)\, \frac{{d\tilde{t}}}{{ \tilde{\tau} }}  ~ = ~ c \, {\tilde{w}}_{0} = \hbar \, {\tilde{\omega}}_{0}  ,
\ee
\noindent  where the value of $n$ in the upper integration limit $n \, \tilde{\tau} \gg \tilde{\tau}$ depends on how well we need to define the photon.  In a system free of gravitational field, we may write, as in Eq.\,(37) and (38), that
\be
\mathbf{S} = \int_0^{\delta \tilde{x}}\!\! c \,\frac{{ {d \tilde{w}(\tilde{x})} }}{{ d{\tilde{x}} }}\, d\tilde{x} = \int_0^{\delta \tilde{x}} c \, {\tilde{w}_{0}}\, \, {\rm{exp}}\left(-\frac{{ \tilde{x} }}{{ c \,\tilde{\tau} }}\right)\, \frac{{d\tilde{x}}}{{ c \, \tilde{\tau} }}~ = ~ c \, {\tilde{w}}_{0} = \hbar \, {\tilde{\omega}}_{0}. 
\ee
\noindent  When one photon is emitted per second in the direction of the x-axis, and when we integrate the electromagnetic energy density not only over the length of the photon, but also from $\delta {\tilde{x}}$ to a very distant observer at $\tilde{x} \rightarrow \infty ,$ we get similar to Eq.\,(38) that
\be
 \mathbf{S} =  \int_0^{\infty} \!\!\! c \, \frac{{ d{\tilde{w}(\tilde{x})} }}{{ d{\tilde{x}} }} \, d{\tilde{x}} = \! \int_0^{\infty} \!\!\! c \, {\tilde{w}_{0}} \,\, {\rm{exp}}\left(-\frac{{\tilde{x}}}{{c\tilde{\tau}}}\right) \,\frac{{ d\tilde{x}}}{{ c \, \tilde{\tau} }} = c \, {\tilde{w}_{0}} =  \hbar \, {\tilde{\omega}}_{0}  .
\ee

\vspace{2mm}


\subsection{Photon energy in a reference system in a gravitational field}

We assume that the metric is given by Eq.\,(16).  The calculations are analogous to those in section 3.4.  The energy flux density is analogous to Eq.\,(32), and we get: $c_*w = c\tilde{w}/\varepsilon^2,~ \varepsilon dx=d\tilde{x},~\tau = \tilde{\tau} \, \varepsilon ,$ and $c_* = c/\varepsilon ,$ and when assuming that the metric is constant over the volume of the photon we get similar to Eq.\,(40) that
\be
c_{*} \frac{{ d{w}(x) }}{{ d{x} }}\,dx = \frac {{c}}{{\varepsilon}} \frac{{ d{\tilde{w}(\tilde{x})}/\varepsilon }}{{ d\tilde{x} }} \, d\tilde{x} =  \frac{{c \tilde{w}_{0}}}{{ \varepsilon^2}} \,\, {\rm{exp}}\left(-\frac{{\tilde{x}}}{{c\tilde{\tau}}}\right) \,  \frac{{d\tilde{x}}}{{ c \, \tilde{\tau} }} , 
\ee
\noindent   where  $c_* \, \tau = c \, \tilde{\tau} .~$  When we integrate the energy density over the length of the photon and assume that the metric is constant and $\varepsilon = \varepsilon_{e}$ over the integration distance, we get analogous to Eq.\,(41) that at the point ${\rm{P}}_{e}$ of emission the energy flux is 
\be
\mathbf{S} =  \int \! c_{*} \frac{{ d{w}({x}) }}{{ d{x} }} \, dx = \! \int \! \frac {{c}}{{\varepsilon_{e}}} \frac{{ d\tilde{w}(\tilde{x}) }}{{ {\varepsilon_{e}} d\tilde{x} }} \, d\tilde{x} = \! \int_0^{\delta \tilde{x}} \!\frac{{c \tilde{w}_{0}}}{{\varepsilon_{e}^{2}}} \,\, {\rm{exp}}\left(-\frac{{ \tilde{x}}}{{c\tilde{\tau}}}\right) \,  \frac{{d\tilde{x}}}{{ c \, \tilde{\tau} }} = \frac{{c\tilde{w}_{0}}}{{\varepsilon_{e}^2 }} = \frac{{1}}{{\varepsilon_{e}}} \frac{{ \hbar \, \tilde{\omega}_{0} }}{{\varepsilon_{e} }} .
\ee
\indent  Eqs.\,(43) corresponds to Eqs.\,(32).  In particular, Eq.\,(43) indicates that the Poynting vector consists of two factors: 
\begin{enumerate}
\item   The first factor $1/\varepsilon_{e} = c_*/c $ shows that the gravitational field at the point ${\rm{P}}_{e}$ of emission reduces the rate of emission.  This factor is often called {\it{time dilation.}} 
\item   The second factor shows that at the point ${\rm{P}}_{e}$ of emission the photon's energy, $\hbar \, {\omega}_{0} = \hbar \, {\tilde\omega}_{0}/\varepsilon_{e} ,$ decreases with the gravitational potential.  
\end{enumerate}
\noindent  This is consistent with Einstein's classical theory for the gravitational redshift.

\indent  Additionally, Einstein assumed that once emitted the photon's frequency would not change as the photon moves from the Sun to the Earth.  Photons frequency was redshifted when emitted in the Sun, as shown by Eq.\,(43), and would be redshifted by the same amount when observed on Earth.  Therefore, according to Einstein we have at the point ${\rm{P}}_{o}$ of observation that
\be
\mathbf{S}  = \frac{{1}}{{\varepsilon_e}} \, \frac{{ \hbar \, \tilde{\omega}_{0} }}{{ \varepsilon_e }}  ,
\ee
\noindent  where ${\varepsilon}_{e} $ is the value of Eq.\,(10) at the point of emission in the Sun.  Einstein argued that in accordance with classical theory, equally many crests of waves had to arrive on Earth as leave the Sun [8].  Therefore, photon's frequency is a constant of motion as the photon travels from the Sun to the Earth and equal to the frequency at the point of emission.  As shown in the introduction, these classical physics assumptions and argumentation by Einstein are impermissible in quantum mechanics and may therefore lead to conflict with experiments.

\indent  We extend the integration over the photon field embedded in a gravitational field from $0 {\rightarrow} {\delta \tilde{x}}$ to $0 {\rightarrow} {\tilde{x}_{o}}$ , where ${\tilde{x}_{o}}$ is the distance to the point ${\rm{P}}_{o}$ of observation, which is assumed far away from the point of emission.  We get that the energy flux density $\textbf{S}$ at the point of observation is
\be
\begin{array}{l}
 \displaystyle \mathbf{S}= \int_0^{\delta \tilde{x}} \! \frac{{c}}{{\varepsilon_{e}}} \frac{{ \tilde{w}_{0} }}{{ \varepsilon}} \, {\rm{exp}}\left(-\frac{{ \tilde{x} }}{{ c \, \tilde{\tau} }}\right) \, \frac{{d\tilde{x}}}{{ c \, \tilde{\tau} }} + \int_{\delta \tilde{x}}^{\tilde{x_{o}}} \! \frac{{c}}{{\varepsilon_{e}}} \frac{{ \tilde{w}_{0} }}{{ \varepsilon}} \, {\rm{exp}}\left(-\frac{{ \tilde{x} }}{{ c \, \tilde{\tau} }}\right) \, \frac{{d\tilde{x}}}{{ c \, \tilde{\tau} }}\\ \quad \\
\displaystyle \quad = \frac{{c \,\tilde{w}_{0}}}{{\varepsilon_{e}^2 }} + \left[ \frac{{1}}{{\varepsilon_{e}}} \frac{{c \,\tilde{w}_{0}}}{{\varepsilon_{o} }} -  \frac{{c \,\tilde{w}_{0}}}{{\varepsilon_{e}^2 }}, \right]= \frac{{1}}{{\varepsilon_{e}}} \frac{{ c \,\tilde{w}_{0} }}{{\varepsilon_{o} }}  = \frac{{1}}{{\varepsilon_{e} }} \, \frac{{ \hbar \, \tilde{\omega}_{0} }}{{ \varepsilon_{o} }} ,
  \end{array}
\ee
\indent  This contradicts Eq.\,(44) and Einstein's assumption that the photon's energy flux and frequency are unchanged as the photon moves outwards from the point of emission to the point of observation.

\indent  However, a priori, we don't know how a photon will behave or how it interacts with the gravitational field after it is emitted at a location, ${\rm{P}}_e,$ close to a star, where $\varepsilon_{e} > 1 .~$  Only experiments can tell us what is right.  This was clear to Einstein.  As shown by Brynjolfsson [1] (see in particular subsections 5.6.1 to 5.6.3 of that source), the interpretation of the solar redshift experiments in light of the plasma redshift theory shows that photons lose their gravitational redshift in accordance with $\hbar \omega_{0} = h\tilde {\omega}_{0} / \varepsilon_{o} $ as they move from the Sun to the Earth.  The solar redshift experiments are thus consistent with Eq.\,(45) and contradict Eq.\,(44).

\indent  Einstein's assumptions may sometimes appear right.  For example, in the experiments by Pound et al. [2-5], the photons did not have time enough to change their frequency, because the height difference was small compared with the photon's length, and because the photon's energy difference at the emitter and absorber is very small.  In the microwave experiments [6,\,7], the microwave photons are very long and reach from Sun to beyond the Earth.  According to quantum mechanics, the microwave photons in these experiments do not have enough time for exchanging energy quanta with the gravitational field during the time of flight from the emitter to the observer.  In these cases we are in the long-wavelength limit, or in the domain of classical physics, which requires that the rate of waves arriving at the detector equals the rate of waves emitted.  The photons' redshifts in these experiments are consistent with Eq.\,(44), because the experimental designs do not make it possible for us to observe quantum mechanical effects.  (If without my knowledge you drop a lead ball on my big toe and look for my reaction within 0.01 second, you will see no reaction, because it takes more than 0.02 seconds for my brain and body to react.  If you observe my reaction 1 second later, you will see my reaction.) 

\indent  In the solar spectrum, most of the photons are only a few meters.  The metric given by Eq.\,(16) can then be considered constant over the volume of the photon.  Most of the photon would have ample time to interact with the gravitational field and exchange energy during their 8 minutes time of flight from the Sun to the Earth.  During this relatively long interaction time, it is possible that the photon's frequency (energy) is not a constant of motion
 
\indent For understanding this better, let us in the next section consider the forces acting on the photon.


\subsection{Forces acting on the photon}

For calculating the components of the electromagnetic four-force density, $f_i ,$ acting on the photon, we need to know electromagnetic energy-momentum tensor $S_i^k ,$ which is given by Eq.\,(30), the metric tensor $g_{ik} ,$ which is given by Eq.\,(16), and the contravariant tensor $S^{kl} ,$ which is defined as 
\be
S^{kl} = g^{ki} \, S_i^l \,\, , 
\ee
\noindent  where the $g^{ki}$ is given by Eq.\,(17).  With the notations used in Eq.\,(30), we get
\be
S^{kl}  =  \left\{ {\begin{array}{*{20}c}
   {\tilde w /\varepsilon^2} & 0 & 0 & { {\varepsilon}^2 \, {\tilde w} }  \\
   0 & 0 & 0 & 0  \\
   0 & 0 & 0 & 0  \\
   {\tilde w/{\varepsilon}^2} & 0 & 0 & { \varepsilon^2 \, \tilde w}  
\end{array}} \right\} \, , 
\ee
\noindent  As Eq.\,(10.304) by M{\o}ller [16] shows, the components of the four-force density $f_i $ are then given by
\be
f_i  = {\rm{div}}_{\rm{i}} \left\{ {S_i^k } \right\} = \frac{1}{{\sqrt {\left| g \right|} }}\frac{{\partial \left( {\sqrt {\left| g \right|} S_i^k } \right)}}{{\partial x^k }} - \frac{1}{2}\frac{{\partial g_{kl} }}{{\partial x^i }}S^{kl} \, {\rm{,}}
\ee
\noindent where $ {\left| g \right|} = {\varepsilon}^4 $ is the numerical value of the determinant of $g_{kl} .$ 

\indent The four-force density in Eq.\,(48) has an opposite sign to the four-force density in Eq.\,(10.304) by M{\o}ller [16].  In Eq.\,(10.304) by M{\o}ller, the electromagnetic field acts on a charge.  In Eq.\,(48) our focus is on the force on the electromagnetic photon field as the photon leaves the atomic charges (we think of the photon being emitted from an atom in accordance with Eqs.\,(34) to (37)). 

\indent  In accordance with Eq.\,(42), we set 
${\partial {w}}/\partial {x} =  {\tilde{w}_{0}}/(c \, \tilde{\tau} \varepsilon) \, {\rm{exp}}\left(-x/(c\tilde{\tau}) \right) .~$  The metric given by Eq.\,(16) does not depend on time.  The four-force density $f_1$ in x-direction (radial direction) is
\be
f_1 = \frac{{1}}{{\varepsilon^2 }} \frac{{\partial \left( \varepsilon^2 \, {\tilde w(\tilde{x}) } \right)}}{{\partial x }}- \frac{{1}}{{2}} \, \frac{{\partial \left(\varepsilon^2 \right)}}{{\partial x}} \,  \tilde w /\varepsilon^2 \, + \frac{{1}}{{2}} \, \frac{{\partial \left( 1 / \varepsilon^2 \right)}}{{\partial x}} \, \varepsilon^2 \, \tilde w  = \frac{{ \partial{\tilde{w}(\tilde{x})} }}{{\partial{x}}}  =  \frac{{ \tilde{w}_{0} }}{{ c \, \tilde{\tau} }} \, {\rm{exp}}\left(-\frac{{\tilde{x} }}{{ c\tilde{\tau} }}\right) \, .
\ee
\indent  As shown by M{\o}ller (see section 10.10 of that source), the corresponding coordinate force density embedded in a gravitational field is $\mathbf{f}_x = f_i/{\varepsilon} .~$  The metric given by Eq.\,(16) is designed to eliminate the gravitational field.  The coordinate force derived from Eq.\,(48) does not include therefore any gravitational-field force.  The coordinate force derived from Eq.\,(48) includes thus only non-gravitational forces.  In this specific case, it includes only the force of the electrical charges pushing the photons outward from the atom (or the photon emitting system).

\indent  We normalize the energy flux to one photon per second as seen by a distant observer at a point free of gravitational field.  We integrate the coordinate force density $\mathbf{f}_x = f_1/{\varepsilon} $ (see section 10.10 of [16]) over the photon field from $0 {\rightarrow} {\delta \tilde{x}}.$  This integration along the extent of the photon emitted at $x = 0$ is made assuming that $\varepsilon_{e}$ is a constant.  We then divide by $\delta \tilde{x}$ and get then the average total coordinate force, $({\mathbf{f}}_x)_{\rm{av}} ,$  pushing the photon outwards.  We thus have that

\be
({\mathbf{f}}_{\delta \tilde{x}})_{\rm{av}} =\frac{{ 1 }}{{ \delta \tilde{x} }}\int_0^{\delta \tilde{x}}  \frac{{ \tilde{w}_{0} }}{{ \varepsilon_{e} }} \, {\rm{exp}}\left(-\frac{{\tilde{x} }}{{ c\tilde{\tau} }}\right)\, \frac{{d\tilde{x}}}{{c \, \tilde{\tau}}} = \frac{{ {\tilde{w}}_{0} }}{{\varepsilon_{e}\, \delta \tilde{x} }} = \frac{{\hbar \, \tilde{\omega}_{0} }}{{ c \, \varepsilon_{e}\, \delta \tilde{x} }} = \frac{{ 1 }}{{ \varepsilon_{e} }} \frac{{ \hbar \, {\tilde{\omega}}_{0} }}{{ c_* {\varepsilon_{e}}\, \delta \tilde{x} }} \, ,
\ee 
\noindent  This force pushes the photon in the direction of x-axis at the point of emission.  When we multiply this average coordinate force, $({\mathbf{f}}_x)_{\rm{av}} ,$ by the distance, $\, \delta \tilde{x}$, we get that the average energy density at the distance $\, \delta \tilde{x}$ is equal to 
\be
{w}_{\delta \tilde{x}} = \frac{{ 1 }}{{ \varepsilon_{e} }} \frac{{ \hbar \, {\tilde{\omega}}_{0} }}{{ c_* {\varepsilon_{e}} }} = \frac{{ \hbar \, {\tilde{\omega}}_{0} }}{{ c\,{\varepsilon_{e}} }}\, .
\ee

\noindent  We multiply this by the velocity of light, which is $c_* ,$ as seen by a distant observer at a point free of gravitationa field.  We get then that the integrated energy flux at $\, \delta \tilde{x}$ is equal to
\be
 {\mathbf{S}}_{\delta \tilde{x}} = \frac{{c_*}}{{c}} \, \frac{{ \hbar \, {\tilde{\omega}}_{0} }}{{ \varepsilon_{e} }} = \frac{{1}}{{\varepsilon_{e}}} \, \frac{{ \hbar \, {\tilde{\omega}}_{0} }}{{ \varepsilon_{e} }}\, ,  
\ee
where, as shown above, the first factor is caused by time dilation, while the second factor is the actual photon energy at ${\delta \tilde{x}}$ close to the point of emission.

\indent  Beyond ${\delta \tilde{x}} ,$ we have that the coordinate force times a small distance $d x$ is equal to  $\left(\partial {w}_{\delta \tilde{x}}  /\partial{x}\right) d x = \left( \,\partial\left(\hbar \, {\tilde{\omega}}_{0}/(c \varepsilon) \right)/\partial{\tilde{x}}\right) d {\tilde{x}} .$  The average force over the distance $x_o - {\delta \tilde{x}} = \Delta {\tilde{x}_o} $ is

\be 
({\mathbf{f}}_{x_o})_{\rm{av}} = \frac{{ 1 }}{{\Delta {\tilde{x}_o} }}\, \int_{\delta \tilde{x}}^{\tilde{x}_{o}}  \frac{{ \partial\left(\hbar \, {\tilde{\omega}}_{0}/(c \varepsilon) \right) }}{{ \partial{\tilde{x}} }} \, d \tilde{x} =  \frac{{ \hbar \, {\tilde{\omega}}_{0} }}{{ c \, {\varepsilon_{o}}\, \Delta {\tilde{x}_o} }} -  \frac{{ \hbar \, {\tilde{\omega}}_{0} }}{{ c \, {\varepsilon_{e}}\, \Delta {\tilde{x}_o} }} \,\, .
\ee 
\noindent  This shows that the average of the non-gravitational forces acting on the photon from ${\delta \tilde{x}}$ to ${\tilde{x}_{o}}$ is not a zero.  When multiplying Eq.\,(53) by the distance $\Delta {\tilde{x}_o}$, we get the energy density increment, $\Delta w_{x_{o}},$ over the distance  ${\delta \tilde{x}}$ to ${\tilde{x}_{o}}.~$  If to $\Delta w_{x_{o}}$ we add the energy density given by Eq.\,(51), we get
\be
w_{o} = {w}_{\delta \tilde{x}} + \Delta w_{x_{o}} =  \frac{{ 1 }}{{ \varepsilon_{e} }} \frac{{ \hbar \, {\tilde{\omega}}_{0} }}{{ c_* {\varepsilon_{e}} }} + \frac{{ \hbar \, {\tilde{\omega}}_{0} }}{{ c \, {\varepsilon_{o}}\,  }} -  \frac{{ \hbar \, {\tilde{\omega}}_{0} }}{{ c \, {\varepsilon_{e}}\,  }} = \frac{{ 1 }}{{ \varepsilon_{o} }} \frac{{ \hbar \, {\tilde{\omega}}_{0} }}{{ c }} \, .
\ee
\noindent  It is seen that the energy density increases outwards from ${\delta \tilde{x}}.~$  The corresponding energy flux at the point $\rm{P}_o$ of observation is obtained by multiplying the right side of Eq.\,(54) by the outward rate or velocity at the point of emission $c_{e} = c/\varepsilon_{e} .~$  We get then
\be
S_{o} = \frac{{ 1 }}{{ \varepsilon_{e} }} \frac{{ \hbar \, {\tilde{\omega}}_{0} }}{{ \varepsilon_{o} }}\, .
\ee
\indent  This equation is identical to Eq.\,(45) and shows that the photon's energy increases from the point of emission close to the star to the point ${\rm{P}}_{o}$ of observation far away from the star.  This contradicts Eq.\,(44) and Einstein's assumptions that the photon's energy flux and frequency do not change as the photon moves outwards from the point of emission to the point of observation.

\indent  The Poynting vector in Eq.\,(55) increases outwards proportional to the velocity of light, $c_{o} = c/\varepsilon_{o} ,$ at the point of observation.  The force, which increases the speed of light as seen by a distant observer, is caused by metric in Eq.\,(16).  In classical GTR, this fictitious force counterbalances the conventional classical gravitational attraction of the photons in a local system of reference.  Therefore, if as Einstein surmised, the photon is gravitationally attracted in the local system of reference by a gravitational coordinate force $F_{i} = - m_{i} G = - m_{g} G  $, where $m_g$ is the gravitational coordinate mass, which is surmised to be equal to the inertial coordinate mass of the photon, and $G$ the coordinate acceleration (see Eq.\,(10.73) and (10.26) of [16]), then the fictitious force, which is repulsive and equal to $ F'_{i} = m_{i} G = - F_{i},$ will balance the attraction.  Therefore, the photons total energy, the Hamiltonian $H = \hbar \, {\tilde{\omega}}_{0}$ is a constant of motion as Einstein concluded [16] (see in particular Eq.\,(10.190) of that source).  The coordinate frequency, ${\tilde{\nu}_0} = {\tilde{\omega}}_{0}/(2\pi) ,$ is then also a constant of motion. 

\indent  However, if the photon is not attracted by the gravitational field in the local system of reference, the fictitious force, $ F'_{i} = m_{i} G =  - F_{i} ,$ created by the metric in Eq.\,(16) will still be repulsive and will increase the photon's energy as seen by a distant observer.  This fictitious force is the reason for the increase in the Poynting vector and the frequency in Eq.\,(55) as the photon moves outwards.


\subsection{Does the equivalence principle apply to static electromagnetic fields?}

We saw above that the conventional quantum theory is consistent with the finding that the photons, as indicated by experiments of the redshift of Fraunhofer lines in the Sun, are weightless in the conventional local system of reference.

\indent  The question then arises about the static or quasi-static classical electromagnetic fields.  Could it be that also more generally, the electromagnetic fields of electrons and protons are weightless?  For this let us first consider if the theory can give any hint.

\indent  The theory stipulated in section 3.3 applies to all electromagnetic fields.  Due to the invariance of the four dimensional volume, we have in the static gravitational field with metric given by Eq.\,(16) that a piece of volume, $\delta V = \varepsilon^3 \delta{x}\,\delta{y}\,\delta{z} = \delta {\tilde{V}},$ is an invariant (see Eqs.\,(10.237) and (10.238) of [16]).  Therefore, the increase in the energy density given by Eq.\,(31) represents also the electromagnetic energy increase in a piece of a volume that moves outwards with the field.  The energy of the electromagnetic field of an electron or a charged nucleus that moves outwards in the gravitational field would then increase as the charged particles move outwards.  This is consistent with the observations, which show that when we bring an atom from the Sun to the Earth, the atomic frequencies are blue shifted.

\indent  The classical physicist, assuming Einstein's equivalence principle, assumes therefore that a gravitational force attracts the electromagnetic field energy of charged particles.  The classical physicist assumes that when the charged particles move outwards, it is because non-gravitational forces lift the charged particles out of the gravitational field and increase their gravitational potential energy.  Therefore, assuming that a charged particle moves slowly outwards so that its kinetic energy is unchanged, $E_{\rm{kin}} \approx 0,$ its potential energy, $V,$ as well as its total energy, the Hamiltonian $H,$ must increase as seen by a local observer.  $H = V + E_{\rm{kin}} \approx V ,$ and therefore $h\nu = H \approx V $ increases as the charged particle moves outwards.

\indent  Another physicist might not be willing to assume Einstein's equivalence principle.  He may assume that the coulomb field of the electron is weightless, and therefore that the gravitational mass, $m_g ,$ is less than the inertial mass $m_i .~$  He assumes that the electromagnetic field surrounding the charged particle is weightless, just like that of the photon.  A distant observer will in this case see the electromagnetic energy in a piece of volume enclosing the charged particle increase outwards because of fictitious forces in the metric given by Eq.\,(16) push the electromagnetic field surrounding the charged particle outwards.

\indent   As in the case of photons, only quantum mechanically valid experiments can resolve which assumption is right.  When analyzing the coulomb field of the nucleus and the electrons, we find that the ratio, $m_{iq}/m_{iA},$ of the inertial mass, $m_{iq},$ of the coulomb fields and the inertial mass, $m_{iA},$ of the entire atom is greater than about 0.001.  We have also that the inertial mass, $m_{iqn},$ of the coulomb field of the nucleus is significantly greater than that, $m_{iqe},$ of the electrons.  The ratio, $m_{iqn}/m_{iA},$ is roughly proportional to, $Z^{2/3},$ where $Z$ is the atomic number.  When we vary the atomic number of the materials by about a factor of two, we should be able to detect the variation in the weight of the electromagnetic field with the atomic number in experiments that are sensitive to a ratio of $m_{iq}/m_{iA} \approx 0.001 .~$  We find that the many free fall experiments, including those carried out by Adelberger et al. [12] and Su et al. [13] are quantum mechanically valid experiments due to the slow oscillations of the test bodies.  These experiments show that the inertial mass is equal to the gravitational mass within an accuracy better than $10^{-11}.~$  Although the energy fractions of the coulomb fields of the nuclei and the electrons are only on the order of 0.001, the weightlessness of the fields would have been detected easily.  We conclude therefore that the electromagnetic fields of charged nuclei and electrons have weights equal to the corresponding inertial masses, or $m_{gq} = m_{iq}.~$


\section{Conclusions}

The experimental finding that photons gravitational redshift in Sun is reversed when the photons travel from Sun to Earth [1] is found to be consistent with conventional quantum theory of physics.  The finding is consistent with weightlessness of the photons relative to a local observer in a gravitational field.  In a reference system of a distant observer, the photons appear as being repelled with a force numerically equal to but opposite to the gravitational attraction.  This finding does not contradict the many experiments, which have been interpreted incorrectly as proving the weight of photon, because the experiments in fact were designed such that the weightlessness could not been observed.  The experiments, which were assumed to prove the weight of the photons, are in the domain of classical physics.  In these classical physics experiments, the photons frequency did not have time enough to change or reverse during the short time-of-flight from the point of emission to the point of detection.  For example, in the experiments by Pound and Rebka Jr. [2,\,3] and Pound and Snider [4,\,5] with 14.4 keV photons from ${\rm{Fe}}^{57}$ the height difference was only 22.5 m, while the minimum height difference required for detecting a reversal of the gravitational redshift on the Earth is about 700 m.  However, during the time-of-flight of the photons from Sun to Earth, the photons have ample time to change their frequency and adjust to the increasing gravitational potential.  The solar redshift experiments are therefore in the domain of quantum mechanics and the reversal of their gravitational redshifts is in fact easily observed when we take into account the plasma redshift of photons, as shown in reference [1].  The quantum mechanical photons behave in some respect like the atoms, which reverse their gravitational redshift frequencies when they are moved from the Sun to the Earth.  In the case of atoms, however, we transfer the energy to the atom when we (or the environment) lift it (to overcome the gravitational attraction) from Sun to Earth.  The corresponding energy transfer is the cause of the reversal of the gravitational redshift of the atomic frequencies.  In case of photons, on the other hand, the gravitational repulsion, which pushes the photons outward as seen by a distant observer, is the cause of the reversal of the gravitational redshift.

\indent  This affects tremendously our cosmological perspective.  For example, it follows that there are no "black holes", because a prerequisite for a "black hole" is the assumption that light is sucked into the "black hole" and cannot escape.  In contrast, the photons can now escape the "black holes".  As shown in section 6 of [1], there is then nothing to prevent us from assuming that under the very high pressure close to the "brink of a black hole limit" matter is annihilated and transformed into photons, which can escape and reform or recreate matter at a distance from the "brink of a black hole limit".  Physicists have seen such annihilation and recreation of matter for many years in the laboratories.  It seems only natural to assume that such processes can also take place at the "brink of a black hole limit" in nature.  In section 6 of [1], we point to several observations, such as the intense positron-electron annihilation radiation seen at the center of the Milky Way.  Also the very high flux of hydrogen moving away from the center of our Milky Way is a strong indication of a recreation of hydrogen close to the center of our Milky Way.  As shown in reference [1] it is therefore possible and even reasonable to assume that the universe can renew itself forever close to "objects at the brink of a black hole limit".  In reference [1], we have shown how the plasma redshift explains the cosmological redshift, the cosmic microwave background, and the cosmic X-ray background.  We have the no need for Einstein's $\Lambda-$term, and it is possible that the universe is quasi-static, and that the matter can be renewed continuously for ever, see reference [1].

\indent  The weightlessness of photons does not conflict with the observed weight of the electromagnetic fields of electrons and nuclei in the free-fall experiments, such as those in references [12,\,13].  That is, the equivalence principle appears valid for electromagnetic fields with the exception of those of photons.  Photons therefore distinguish themselves not only by their exceptional velocity and zero restmass but also by their weightlessness in a local system of reference.


\end{document}